# Origins of colors variability among C-cluster main-belt asteroids

Pierre Beck[1,2], Olivier Poch[1]

[1]Institut de Planetologie et d'Astrophysique de Grenoble, UGA-CNRS

[2]Institut Universitaire de France, Paris, France

Abstract: The Sloan Digital Sky Survey provides colors for more than 100 000 moving objects, among which around 10 000 have albedos determined. Here we combined colors and albedo in order to perform a cluster analysis on the small bodies population, and identify a C-cluster, a group of asteroid related to C-type as defined in earlier work. Members of this C-cluster are in fair agreement with the color boundaries of B and C-type defined in DeMeo and Carry (2013). We then compare colors of C-cluster asteroids to those of carbonaceous chondrites powders, while taking into account the effect of phase angle. We show that only CM chondrites have colors in the range of C-cluster asteroids, CO, CR and CV chondrites being significantly redder. Also, CM chondrites powders are on average slightly redder than the average C-cluster. The colors of C-cluster members are further investigated by looking at color variations as a function of asteroid diameter. We observe that the visible slope becomes bluer with decreasing asteroids diameter, and a transition seems to be present around 20 km. We discuss the origin of this variation and, if not related to a bias in the dataset - analysis, we conclude that it is related to the surface texture of the objects, smaller objects being covered by rocks, while larger objects are covered by a particulate surface. The blueing is interpreted by an increased contribution of the first reflection in the case of rock-dominated surfaces, which can scatter light in a Rayleigh-like manner. We do not have unambiguous evidence of space weathering within the C-cluster based on this analysis, however the generally bluer nature of C-cluster objects compared to CM chondrites could be to some extent related to space weathering.

1. Introduction

The present-day Solar System can be classified as a debris disk. It hosts a large population of objects referred as small bodies, that are the leftovers of planet formation, and offer an opportunity to access the mineralogy and composition of the materials that were incorporated in terrestrial planets and giant planetary cores. An important reservoir of small bodies is located between Mars and Jupiter, main-belt asteroids (MBAs). Among MBAs, a major fraction belongs to the so-called C-complex (DeMeo et al., 2009), being identified by a low-albedo and the lack of strong anhydrous silicate absorption features in their visible and near-infrared spectra. They have been historically linked to carbonaceous chondrites, a particular type of meteorites belonging to a distinct chemical reservoir when compared to the other types of chondrites (Trinquier et al. 2007, Warren et al. 2011).

The construction of classification schemes for small bodies has enabled to assess their compositional diversity, and to map their geographical distribution across the Solar System (Gradie and Tedesco, 1982; DeMeo and Carry, 2014). The classification schemes are multiple and evolving with time, as the number and type of observations increase, and new classification methods emerge. At the same time, debris of asteroids are naturally delivered to Earth in large amounts, meteorites, which provide an opportunity to anchor and understand the spectral diversity of small bodies in term of composition. Because only a few compounds have absorptions in the visible and near infrared spectral range, the meteorite approach is key: as an example we would not know that the silicate feature observed on S-type asteroids is related to the presence of chondrules, and the spectra of S-type could be matched by an olivine- and pyroxene-rich rock, the mantle of a terrestrial planet for example.

The meteorite approach has also several limitations. First, the Earth atmosphere filters out most of the fine-grained material, and the extra-terrestrial material collections is immensely biased toward rocky material. Second, laboratory measurements on meteorite are not done in the same geometrical conditions as small bodies observations. They are typically measured at 30° phase angle while main-belt asteroids observation have phase angle ranging from 0 to 30°. In addition, ground-based observations are integrated over a hemisphere, and reflectance is not controlled by phase angle only but also incidence and emergence (Potin et al., 2019). Approaches have been developed to account for these effects in observation of Near-Earth Asteroids (Binzel et al., 2015). Last but not least, airless bodies are exposed to a number of specific processes that may alter the physical and chemical properties of their surfaces and these processes need to be emulated in the laboratory to understand how they impact the meteorite/asteroid spectral connection.

A first process that occurs on asteroid surfaces is macroscopic impacts. The heat deposited by hypervelocity collisions can lead to a transient increase of temperature and more prolonged heating in the case of burial below ejecta blankets. In addition to macroscopic impact, micrometeorites impacts can also alter the surface composition by leading to the deposition of small but optically modifying layers of vapor condensate, or the production of agglutinates (Hapke, 2001). Last, the exposure to irradiation by the solar wind and galactic cosmic rays can also change the optical properties of silicate-rich surfaces. A review on the understanding of the processes that may alter the optical properties of airless bodies, space weathering, can be found in Pieters and Noble (2016).

Space weathering is now largely accepted as the process that explains the differences between S-type asteroid spectra and reflectance spectra of ordinary chondrites. Space weathering on S-type is likely the results of faster changes due to solar wind exposure, and slower modification related to micrometeorites impacts (Vernazza et al., 2009). In the case of C-types asteroids, important progresses have been made recently on simulating the effects in the laboratory (Lantz et al. 2017, Brunetto et al. 2020, Matsuoka et al. 2020, Thompson, et al., 2020), in the framework of sample-return missions to near-Earth asteroids. The effects appear to be less pronounced than in the case of ordinary chondrites, with generally a bluing of dark sample with increasing irradiation dose, whether in the case of solar wind simulation or micrometeorite impact simulations (Lantz et al. 2017; Brunetto et al. 2020; Matsuoka et al. 2020; Thompson, et al., 2020). Hints of space-weathering have been provided by the Small Carry-on Impactor experiment on board Hayabusa2, which revealed a small shift of the 2.7 μm metal-OH absorption in subsurface material compared to typical surface spectra (Kitazato et al., 2021). Still, unambiguous evidence of space weathering effects on C-type asteroids in the visible range are absent at present, and while lunar-style space-weathering products have been identified in ordinary chondrite regolith breccias (Noble et al., 2010) they have not been observed so far in carbonaceous chondrite regolith breccias, to the best of our knowledge.

In the present work, we investigate the variability of colors of C-type main-belt asteroids based on the Sloan Digital Sky Survey dataset (SDSS) with the intent to compare to laboratory spectra of carbonaceous chondrites, discuss the C-type carbonaceous chondrites relations, and search for evidences of space-weathering. The data processing of the SDSS dataset is in essence identical to that presented in DeMeo and Carry (2013). C-complex asteroid are then extracted from a cluster analysis combining colors and albedo. This work builds on a recently developed approach to assess the impact of observation geometry on meteorites spectra (Beck et al., 2021). We show that C-type asteroids are overall bluer than carbonaceous chondrites, which may suggest that space-weathering is generally affecting the surface of C-type main-belt asteroids. However, we find that large C-type asteroids (D>20 km) are redder than small C-type asteroids (D<20 km). This difference could be related to a sampling artifact of the SDSS dataset, biased toward smaller objects. If not, we suggest that this overall blue nature of smaller C-type is related to the presence of coarser-grained material.

2. Method

2.1 Colors and albedo

This work uses data from the 4th release of the Sloan Digital Sky Survey (SDSS). The data processing was done following the procedure developed in DeMeo and Carry (2013). Chiefly, SDSS offers a color filter survey of about 400 000 Solar System moving objects. The colors used for this work are g' (0.4686 μm), r' (0.6166 μm), i' (0.7480 μm) and z' (0.8932 μm) while the u' band was excluded for the reason described in DeMeo and Carry (2013). We used color specific magnitude thresholds as defined in DeMeo and Carry (2013) to exclude the faintest objects from the catalogue. Objects with too high photometric uncertainties were also excluded from the dataset. For each object, relative reflectance (normalized to the g' filter) is calculated using:

$$Rf = 10^{0.4*\{(M_f - M_g) - (M_{f,\odot} - M_{g,\odot})\}}$$

Where $M_f$ and $M_{f,\odot}$ are the magnitudes in a filter for the object and for the Sun respectively. Solar colors are taken from Holmberg et al. (2006) following DeMeo and Carry (2013). Albedos and diameters used in this study were taken from the WISE catalog (Masiero et al., 2011).

## 2.2 Cluster analysis

Cluster analysis was performed using a routine available within the IDL environment. Three components were used for the cluster analysis. The first two are the *g'r'i'* spectral slope (referred to as *gri* slope in the rest of the manuscript, %/100 nm) and *z-i* magnitude difference. These two parameters were shown to display a large spread in DeMeo and Carry (2013) analysis and were used to map the compositional structure of the asteroid population. The *gri* slope corresponds to the visible slope of each object while *z-i* is an indication of the presence of an Fe-related absorption band.

Because historically C-type asteroids have been identified as being dark objects, albedo was added as a third component to the cluster analysis. Before running the algorithm, the average of each 3 parameters were calculated and subtracted. Then the standard deviation was calculated and each of the 3 parameters was divided by its standard deviation. From there, cluster analysis was performed with equal weights given to each of the three parameters. Data with the following parameters were filtered out *gri* slope > 30 or < -10 %/100 nm, *z-i* > 0.25 or < -0.75, and albedo > 0.5. This enables a better definition of the clusters by minimizing the impact of strong outliers in the analysis. An arbitrary number of clusters was used in the analysis (n = 9). This number was chosen since it enables to have a cluster definition reminiscent of the *gri* slope and *z-i* boundaries from DeMeo and Carry (2013). It appeared that the position of the "C-type" clusters was not impacted by the number of cluster chosen as long as n > 5.

## 2.3 Laboratory data and phase angle effects

Laboratory spectra were taken from the RELAB database, which provides today the most extensive suite of reflectance spectra of extra-terrestrial materials. The Tagish Lake spectra were measured at IPAG on the same lithologies described in Gilmour et al. (2019).

One of the difficulties of comparing laboratory spectra of meteorite to asteroid observation is the impact of observation geometry, in particular in the case of spectral slope. This work builds on the recent study of Beck et al. (2021a) that investigated the effect of observation geometry on reflectance spectra of a series of about 70 different meteorites. This showed that for dark objects reflectance and slope could be significantly impacted, but that these effects can be accounted for. Building on that work, for each reflectance spectra used here and measured at g = 30°, the same spectra could be calculated for g = 0° (Beck et al., 2021).

3. Definition of the C-type cluster

The work presented here strongly depends on how asteroids are attributed as members of the C-type cluster, which deserves some discussion. Once run, the cluster analysis shows the presence of a large cluster of objects with *gri* slope and *z-i* within the range of the "C-box" defined in DeMeo and Carry (2013) (Figure 1) and overlapping with the B and X-boxes. This cluster will be defined as the C-cluster. The cluster analysis did not reveal the presence of a B-cluster, showing that B-class asteroids do not represent a distinct population of objects. Rather, it appears that the transition from C- to B-type object is a continuous process when considering SDSS colors only. On Figure 1 (top left) the *gri* slope vs *z-i* array is displayed with a threshold applied on albedo values. This reveals that most of members of the C-cluster correspond to a dark population of objects. While B- and C- type cannot be distinguished in our cluster analysis, it enables to separate a C-type from a D-type related cluster (cyan and upper violet clusters in Figure 1 (bottom right) and figure 2.

The clusters identified are presented in Figure 2, and were grouped according to their expected link to taxonomic endmember. Their proper element distribution is also shown in Figure 3. While difficulties may exist in distinguishing C-type from X-type or D-type objects using colors only, the cluster analysis was efficient in separating a cluster overlapping with C and D-type in term of colors but with a significantly higher albedos (Cluster 0) that are more related to S, K or X-complex asteroids based on the proper element distribution. Discussion and investigation of the different cluster identified will be the subject of further work, and we will focus here on the C-cluster.

4. Comparison of C-cluster asteroid colors to carbonaceous chondrites colors

In Figure 4, a comparison is made between colors of C-cluster members and colors calculated for carbonaceous chondrites spectra (in the *gri* slope vs *z-i* space), under standard geometry g=30° or after applying a correction to g=0° (described in section 2). The effect of the phase angle correction is to generally decrease the *gri* slope of studied samples (Figure 4) while *z-i* does not change significantly. The comparison is made with CV-CR-CO (Figure 4 top), CM and heated CM (Figure 4 middle) and Tagish Lake (Figure 4 bottom). Several observations can be made from this diagram. First, carbonaceous chondrites belonging to CV, CR and CO groups (about half of carbonaceous chondrites falls) are significantly redder that C-complex asteroids. This clearly shows that a significant fraction of carbonaceous chondrites does not match the colors of C-type asteroids. Such a conclusion has been reached previously based on spectroscopy (therefore mostly for the largest objects) and based on albedo considerations (Bell 1988, Burbine et al. 2001, 2002; Clark et al. 2009, Eschrig et al., 2021). Because terrestrial weathering is known to affect colors and spectra of meteorites, we investigated the difference between falls and finds in this diagram. In the case of CM chondrites, the meteorite falls studied (Mighei, Murray and Murchison) span almost the range of value for all the CM studied (falls and finds). It appears then than terrestrial weathering does not impact significantly the colors of CM falls (or that it affects both falls and finds). In the case of CR, CV and CO chondrites, which can contain a significant amount of metal and sulfides, there is a clear difference between falls and finds. The *gri*-slope value for falls is between 3 and 12 %/100 nm, while the *gri*-slope value for finds is between 1 and 41 %/100 nm (Figure 4). Such a difference could be attributed to the terrestrial formation of $Fe^{3+}$-rich oxides at the expense of Fe-metal. When comparing the *gri*-slope and *z-i* of CR, CV and CO falls against the boundaries defined by DeMeo and Carry (2013), it appears that the *gri* slope values of these meteorites powders are higher than the average C-cluster, and more in agreement with the X- and K-type boundaries.

In the case of the ungrouped carbonaceous chondrite Tagish Lake, colors of powdered samples clearly fall outside of the C-type cluster defined here, and falls within the X and D boxes defined in DeMeo and Carry (2013) (Fig. 4). This is somehow in agreement with the proposition that Tagish Lake is a fragment of a D-type asteroid, and as discussed in Beck et al. (2021a), more likely a fragment of the interior of a D-type.

Among the suite of carbonaceous chondrites studied here, only CM chondrites appear to correspond to the range of values of C-cluster asteroids. Among those, samples that have been described as heated CM chondrites (Nakamura et al., 2005, Alexander et al., 2012, Quirico et al. 2018) tend to show a more pronounced *gri* slope than non-heated CM chondrites (Fig. 4 middle), while their *z-i* value remains similar. In addition, on average CM chondrites appear to show redder slopes than C-cluster members. This is confirmed by analyzing the distribution of *gri* slope values of CM chondrites and comparison to C-cluster members (Fig. 5).

5. Discussion

5.1 The B-type / C-type continuum

B-types asteroids spectra in the Vis-Nir are defined as "Linear, featureless spectrum over the interval from 0.44 to 0.92 µm, with negative (blue) to flat slope" (Bus and Binzel, 2002). The B-type was identified as a spectral endmember in the DeMeo et al. (2009) taxonomy, where the B-type spectrum resembles C-type, but to which a blue slope has been added. Famous objects classified as B-type includes 2-Pallas, 3200-Phaeton or 335-Roberta. B-type asteroids tend to have spectra showing a broad 1-µm feature and sometimes 3-µm band indicative of hydrous minerals (Yang and Jewitt, 2010), but not always (Takir et al., 2020). The broad 1-µm feature has been interpreted as magnetite (Yang and Jewitt, 2010).

In the color-based classification scheme of DeMeo and Carry (2013), B-type asteroids represent 11.10 % of the mass of the main-belt, and 3.55 wt.% when Pallas is removed. When the cluster analysis was applied to the dataset studied here, there was no cluster identified with negative *gri* slope value and low-albedo, which would correspond to B-type asteroids. Rather, this analysis suggests that there is a continuum of objects between B- and C- type rather than two distinct categories. This is in agreement with the somehow arbitrary definition of the B-type, as being objects bluer than the Sun.

In order to assess the potential "separability" of a C-cluster from a B-cluster, we need to address whether the slope variability found within the C-cluster is true, or related to observational uncertainty and biases. First, the slope variability found is significant, and can hardly be explained by observation uncertainty or phase angle effects. While in some special cases phase angle can change the spectral slope from red to blue (Jost et al., 2017a,b), measurements performed on carbonaceous chondrites showed that phase angle can decrease or increase slope, but not change its sign for the phase angle range of interest (typically 0-30°, Beck et al., 2021a).

Second, error on color magnitudes may transpose into error on the slope determination. We investigated this effect by using average error on color for members of the C-type cluster (g=0.0218, r=0.0171, i=0.0181) and propagating this error on *gri* slope determination using a Monte-Carlo approach. This analysis enables to estimate the uncertainty on *gri* slope for the C-cluster to be +/- 0.934 %/100 nm (1 sigma). This analysis suggests that some of the spread in *gri* slope observed (Figure 1 and 5) is due to uncertainty on color magnitudes, but that the overall variability within the cluster is real and that identification of a B-type cluster may have been possible.

The presence of a continuum of objects with varying slopes can be explained whether by the presence of a continuum of composition between a B and C endmembers, or by a process that continuously modifies the C signature onto a B signature or vice-versa.

5.2 Size dependence of the *gri* slope

In order to identify the origin of the slope variability among the C-cluster, the relation between the *gri* slope was studied as a function of asteroid diameter (Fig. 5). This analysis revealed than when considering objects smaller or larger than 20 km in diameter, there is a difference in the distribution of *gri* slope. The average *gri* slope of large objects is higher by about 1 %/100 nm when compared to smaller objects. This is a small difference but that appears significant. On Figure 5 is also shown the *gri* slope distribution for CM and heated CM chondrites powders, which appears more similar to the larger C-types (>20 km) than the smaller ones.

In order to further investigate the variability of colors with size, the *z-i* and *gri* slope were plotted as a function of diameter, and running median and average were calculated over 50 elements (Figure 6). This shows that both *z-i* and *gri* slope decrease with decreasing diameter. In the case of *gri* slope two regions seem to be present with a slope change around 20 km diameter objects. In line with the observations discussed here, Vernazza et al. (2016) observed a spectral evolution with asteroid diameter in a spectral survey of about 70 Ch/Cgh objects (sub-type of the C-complex). The smaller object analyzed in Vernazza et al. (2016) tend to be bluer than larger ones, as well as to present fainter phyllosilicate absorption bands.

Because the SDSS dataset (and the asteroid belt) contains much less large asteroid than small asteroids, one needs to verify that the difference in *gri* slope observed is not due to a number-related bias. Considering large (>20 km) vs small (<20 km) objects, the average values of *gri* slope seems so be statistically different if the error on the mean (standard deviation divided by the square root of the number) is considered (2.27 %/100 nm ± 0.10, n=470; 1.43 %/100 nm ± 0.03, n=6272). Nevertheless, it cannot be strictly ruled out that some bias exists in the dataset or analysis that would produce the *gri* slope vs diameter trend. We tested phase angle as a possible source of bias, but the range of phase angle are similar between large and small objects, and no correlation was found between phase angle and *gri* slope. Because large and small objects are on average of different magnitude, the difference may be caused by a magnitude dependent bias. But at the moment we do not have evidence for such a bias.

To summarize 5.1 and 5.2, there is a continuous variability of slope within the C-cluster, and smaller members seem to be slightly bluer than larger members. We will suppose that this variation is true, and discuss in the following sections several possible causes for this variability.

5.3 Space weathering

In the case of S-type asteroids, space-weathering effects appear widespread with the exception of small near-earth asteroids whose surface might be refreshed by physical effects (Binzel et al., 2004). In the case of C-type asteroids, considerable efforts have been undertaken in the last years to understand the consequences of high-energy irradiation as well as micrometeorites impact from a laboratory perspective. In the case of ordinary chondrites, containing large crystal (>50 μm) of translucent silicates (olivine & pyroxene), the production of small nano-phase iron particles whether by vapor deposition or *in situ* transformation within the grains lead to an increase of spectral slope, and a decrease of visible reflectance and absorption band contrast (Hapke, 2001). The lack of strong silicate feature on fresh carbonaceous chondrites spectra implies that the spectral effects of space weathering will be different from ordinary chondrites. Another option to investigate space-weathering effects is to observe the spectral variability of asteroid families as a function of dynamical age (Vernazza et al., 2009) or to observe spectral signature of freshly exposed material from asteroids surfaces (by impact of mass wasting processes). In the case of S-type, these two approaches have been used and suggest that the spectral effects observed in the laboratory reproduce well those observed on asteroid surfaces (Chapman, 1996).

Laboratory investigations of space-weathering effects on carbonaceous chondrites have shown behavior quite distinct from S-type, with bluing or reddening of the samples upon ion irradiation (Lantz et al., 2017), and bluing in the case of simulated micrometeorite impacts using pulsed laser (Matsuoka et al., 2020). Spectra from these studies were digitized and used to calculate the impact of space weathering on the *z-i* vs *gri* slope diagram (Fig. 7). Within this array, simulated micrometeorite impact produces a diagonal vector pointing toward lower *z-i* and lower *gri* slope. In the case of ion irradiation studies on Mighei (CM) and Tagish Lake (UCC: ungrouped carbonaceous chondrite) (Lantz et al., 2017), the effect is to induce a bluing without impacting the *z-i* value. Note that in the case of these two samples, there were no absorption around 0.9 µm in the starting material, and the lack of impact on the *z-i* value can be attributed to the absence of feature in the initial sample (a feature that is not present cannot disappear).

Observations of C-type asteroids by the Hayabusa2 (HY2) spacecraft is providing important clues on space-weathering of C-type Near Earth Objects (NEOs). The Small Carry-on Impactor (SCI) experiment onboard HY2 seems to show that space weathering effects are not as pronounced as for C-type asteroids, as suggested by space weathering of dark carbonaceous chondrite (Lantz et al., 2017). It is not known at the time of the writing what the exact value is for the inferred vector for space weathering in the *z-i* vs *gri* slope space as observed remotely for C-type. In the case of CO and CV chondrites, space-weathering tends to redden the spectra, which further excludes a connection between C-type and these groups of carbonaceous chondrites.

A conclusion that can be drawn from laboratory studies of space weathering effects is that the difference in *gri* slope between carbonaceous chondrites powders (CM, 3.8 %/100 nm on average) and C-type asteroids (1.3 %/100 nm on average) (Figure 4 middle) can be explained by space weathering because it is known to induce a shift of *gri* slope by 2 to 4 %/100 nm (Figure 7).

One observation that is not easily attributable to space weathering is the difference in *gri* slope between large and small asteroids. One might expect that being of collisional origin and therefore with a younger surface, small asteroids may be somehow less affected by space weathering than larger objects. This is contrary to the observation of Figure 4 and 5. A possibility is that larger objects may be shielded from some of the solar wind particles, if a remnant of a magnetic field is present. This has been proposed for 4-Vesta (Vernazza, et al., 2006), may be applicable to Ceres (Neuman et al., 2015), but is difficult to imagine for C-type asteroids in the 20-200 km diameter range. Their density is around 1.4 (Carry, 2012) and their shape does not suggest hydrostatic equilibrium which would be expected from differentiation of a metallic core. Recent study of the C-type Daphne appears to reveal a fairly homogeneous structure (Carry et al., 2019). As discussed in the following, other effects may explain the difference in spectral slope between large and small C-type as well as the overall slope variability.

5.4 Rayleigh-like scattering particles

The presence of particles smaller than the wavelength can lead to blue spectra as shown experimentally (Clark et al., 2008; Poch et al., 2016; Sultana et al., 2020), and the presence of such "hyperfine" particles has been proposed to explain the blue areas of the C-type Ceres (Schroder et al., 2021). The size of the individual constituents of the matrix of carbonaceous chondrites is μm to sub-μm as well as the size of the individual constituents of Inter-planetary Dust Particles (IDPs) or cometary dust. Therefore, if the hypothesis that the "non-hydrated" C-types are made of IDP-related material is correct (Vernazza et al., 2015), this would imply that the surface material of many C-types is constituted of dust with sub-μm building blocks. If hyperfine particles are more abundant in the case of the smaller C-type asteroids, this could explain the color difference when compared to larger objects. However, because of the recurrent collisions (micro and macro) occurring on their surfaces, we may expect some degree of compaction of asteroidal surfaces. If initially constituted by hyperfine particles, compaction may suppress a Rayleigh-like behavior and a process is needed to be active in restoring the porosity. This could be sublimation of dust-rich ice that would produce upon sublimation a fluffy residue where Rayleigh-like scattering may occur (Poch et al., 2016; Schroder et al., 2021, Sultana et al., 2020).

However, this scenario would suggest that the smaller and bluer objects have a surface finer-grained or/and with higher porosity, and therefore a lower thermal inertia. As discussed below this is at odd with observations of thermal inertia for asteroids of difference sizes.

5.5 Grain size effect and Rayleigh-like scattering by grain surfaces

Grain size are known to have a significant impact on reflectance spectra of carbonaceous chondrites, and more generally texture (powder, pellet, chip) can affect spectral slope and the depth of absorption features (Ross et al., 1969; Johnson and Fanale, 1973, Cloutis et al., 2018). In order to assess the impact of sample texture, we used reflectance spectra obtained on small fragments of fresh CM chondrites, which were subsequently ground into powders. These data obtained on the Aguas Zarcas CM fall are presented in Beck et al. (2021b), and revealed that CM fragments are much bluer than powders, and show fainter absorption features. The reason for this reddening after grinding may be due to the comminution of red-sloped phases that may become more visible when finely mixed with other minerals (Ross et al., 1969). In other words when grinding a sample, the mixture becomes intimate (proportional in single-scattering albedo space to the surface area of grains) while in the case of chips the mixture is more geographical (proportional to the surface area of the various phases).

Another possibility is than when grinding the sample, and increasing the specific surface area, the interaction with Earth's atmosphere is magnified, producing some oxidation of the sample. This may explain some observation related to the D/H isotopic composition of CM chondrites (Vacher et al., 2020) if hydroxides or oxy-hydroxides are produced by this process. However, observation of a similar slope change for ordinary chondrites that are likely less reactive than carbonaceous chondrites may disqualify this explanation.

Clark et al. (2008) suggested that the blue slope measured for some natural rock samples could be explained by Rayleigh-like scattering. This is our favored hypothesis to explain the blue slope of meteorite rocks or large grained powder (>1 mm) when compared to finer powder. In the case of absorbing compact material, the intensity of the reflected light will be strongly controlled by the first external reflection, at the rock surface. If asperities smaller than the wavelengths are present at the surface of the rock, they may scatter light in a Rayleigh-like manner and produce the increase of reflectivity with decreasing wavelength observed. If the same rock is ground to finer powder, multiple scattering may start to contribute more to the measured reflectance, which will be less sensitive to the first external reflection, and then surface asperities. This is consistent with emission observations of the silicate features from Marchis et al. (2012) that revealed more contrasted emissivity signatures for the larger objects.

When the spectra of raw and powdered carbonaceous chondrites are plotted in the *z-i* vs *gri* slope diagram (Fig. 7), a shift of typically 2 %/μm is found for the *gri* slope. This change is of the order of the difference in box-averaged *gri* slope between small and large C-type (Figure 5). The hypothesis of a variation of grain size with diameter was favored by Vernazza et al. (2016) to explain the spectral variability of Cgh/Ch objects. It is also supported by variation of thermal inertia with asteroid diameter as reported in Delbo et al. (2007). Thermal inertia is obtained through infrared observations of asteroids, and is directly related to heat capacity and thermal conductivity of surface material. Powder surfaces tend to have a low thermal inertia, while large block tend to have a high thermal inertia. The fact that larger objects are covered by material with low thermal inertia, contrary to smaller objects, points to the presence of finer-grained material at the surface of large asteroids.

Radar measurements of the SC/OC ratio of planetary bodies enables to probe a combination of grain size and roughness at a scale of the order of the radar wavelength, therefore several cm. Radar observations of small asteroids are dominated by NEA, while radar observations of large objects are dominated by MBAs. These observations reveal a clear decrease in the SC/OC ratio of small C-type (i.e. NEAs, 0.285 +/- 0.120) compared to large C-type (i.e. MBAs, 0.098 +/- 0.058, all objects observed being larger than 20 km diameter) (Benner et al., 2008). The scarcity of radar observation of C-type objects (total of 42 objects in Benner et al., 2008) does not permit to pinpoint a particular diameter for the transition. Still the very low SC/OC ratio of large C-type has been interpreted by the presence of a fine-grained regolith (not rough or not scattering at the scale of the radar wavelength), while the SC/OC ratio of small objects points toward a rougher or lager grained surface, in line with the conclusion we derive in our manuscript from optical observations.

6. What is the favored process to explain colors within the C-cluster?

Based on the discussion above we have identified three processes that can explain the difference in colors of C-cluster asteroids and carbonaceous chondrites. They are space-weathering, the absence of a regolith on C-type MBAs, and the presence of a porous sublimation lag. All these three processes can explain the C-cluster asteroid vs carbonaceous chondrites difference, but at the moment only one of them can also explain the difference in slope between large and small C-type MBAs: grain-size effect. As discussed above, a surface covered by rocky fragments rather than regolith is likely to present a blue slope, as observed in the case of C-type NEOs. The present analysis suggests that small C-type MBAs host on their surface a significant fraction of rocks rather than a fine-grained material, in agreement with thermal inertia measurements (Delbo et al., 2007).

The transition from redder to bluer C-type MBAs may be related to the transition from primordial objects to fragments, i.e. small family members. Based on the "wavy" nature of the size frequency distribution of MBAs, it has been proposed that D<100 km asteroids represent objects produced by the collisional fragmentation of larger objects (Bottke et al., 2015). Most of the small asteroids from the C-cluster defined here belong to dynamical families and were therefore produced by collisions (Fig. 3, Nesvorny et al., 2015; Delbo et al., 2017).

But being a collisional fragment does not imply that the surface should be devoid of fine-grained material. Many asteroid families are ancient (>100 My), and even small objects may have had time to develop a fine-grained regolith. Even if an asteroid is a rubble-pile, exogenous processes will still act and may lead to the production of surface fines. However, because of the low escape velocity for the smallest asteroids, the preservation of fine-grained material produced by impact will be more difficult. Different style and thickness of regolith are expected to be present on small bodies depending on their size, as modeled by Housen et al. (1979). We propose here that the progressive bluing for smaller objects is related to the presence of more rocky regolith. The "rockier" nature of smaller asteroids may be related to the fact that fine-grained material escape during impacts, or other physical process that may enrich the surface in larger grains material, like the "Brasil-Nut" effect (see review by Murdoch et al. 2015 on physical processes that may occur on micro-gravity environment).

This does not mean that space-weathering does not impact on the colors of C-cluster. While the maxima of the distributions are similar (Figure 5), on average large C-cluster asteroids (D>20 km) have a slightly bluer spectral slope than average CM chondrites powders from our study (2.3 vs 3.6 %/100 nm). This difference could be explained by the presence of a fraction of rocky material at the surface of large C-types as well. However thermal inertia observation of large C-type suggest that their surface is covered by a fine regolith rather than rocks, and D > 50 km C-type objects can be expected to be primordial rather than collisional in origin (Marchis, et al., 2012; Delbo et al., 2017). In that case, the effects of space-weathering may be invoked to explain this difference and both irradiation and micrometeorite provide efficient ways to turn bluer the surface of large C-types when compared to carbonaceous chondrites. These effects are in any case modest when compared to those observed for S-complex objects.

7. Conclusions

In this article, we have presented an analysis of colors of dark asteroids from the SDSS dataset with a focus on "C-type". We found that:

- A cluster analysis based on *gri* slope, *z-i* and albedo enables to identify a cluster that can be assigned to "C-type" with boundaries in the *z-i* vs *gri* slope space close to the ones defined by DeMeo and Carry (2013). The analysis does not identify a B-type cluster, but rather enlightens the presence of a continuum of objects with varying visible slopes.

- CM chondrites are the closest match to the C-cluster identified. Other types of carbonaceous chondrites studied here are not (CO-CV-CR and Tagish Lake). Carbonaceous chondrites of CO, CR and CV groups have colors more similar to X-, K- and L-type as defined in DeMeo and Carry (2013).

- There appears to be a size dependency of the colors of C-cluster members, which merits further investigation and confirmation by other datasets. The smaller objects (D<20 km) appear bluer than larger objects. The preferred explanation is that smaller C-cluster asteroids have surfaces covered by rocks, or large grained material (>1 mm) rather than a fine-grained regolith. This suggest that a size-sorting mechanism may act under micro-gravity environment. Possible mechanisms include escape of finer material during impact of the "Brazil-nut" effect.

- CM chondrites powders are slightly redder than large C-cluster asteroids that are expected to be covered by regolith. This difference may be attributed to high-energy irradiation based on available laboratory data.

**Acknowledgments**: This work was funded the European Research Council under the H2020 framework program/ERC grant agreement no. 771691 (Solarys). Comments by two reviewers greatly improved the manuscript. Additional support by the Progamme National de Planétologie and the Centre National d'Etude Spatiale is acknowledged.

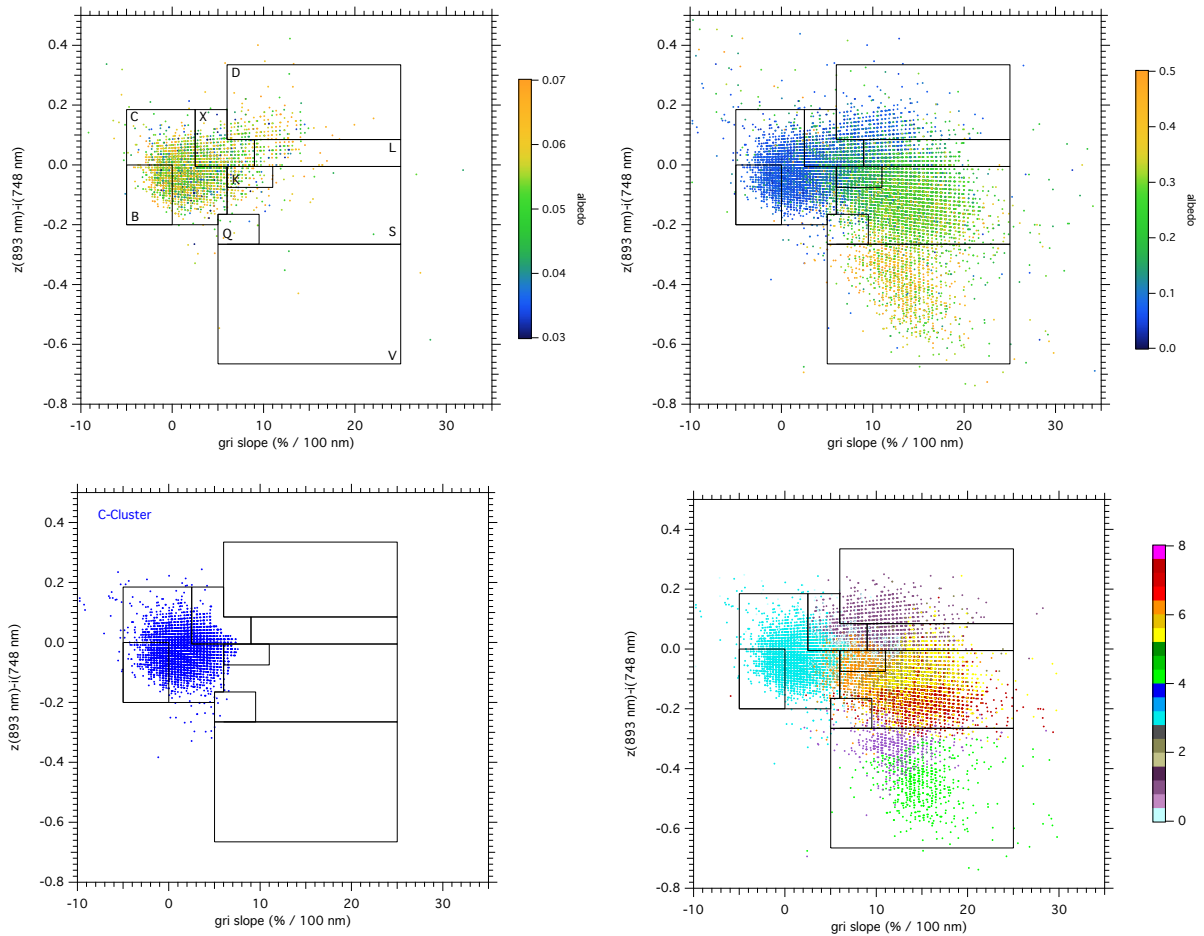

Figure 1: *z-i* vs *gri* slope diagram for the asteroid subset extracted from the SDSS database (see text for details). Top left: All objects with albedo <0.07. Top right: All objects color-coded with albedo value. Bottom right: All clusters identified. Bottom left: objects from the cluster defined as C-type. The boxes correspond to the limits defined by DeMeo and Carry (2013). Gri slope is in %/100 nm.

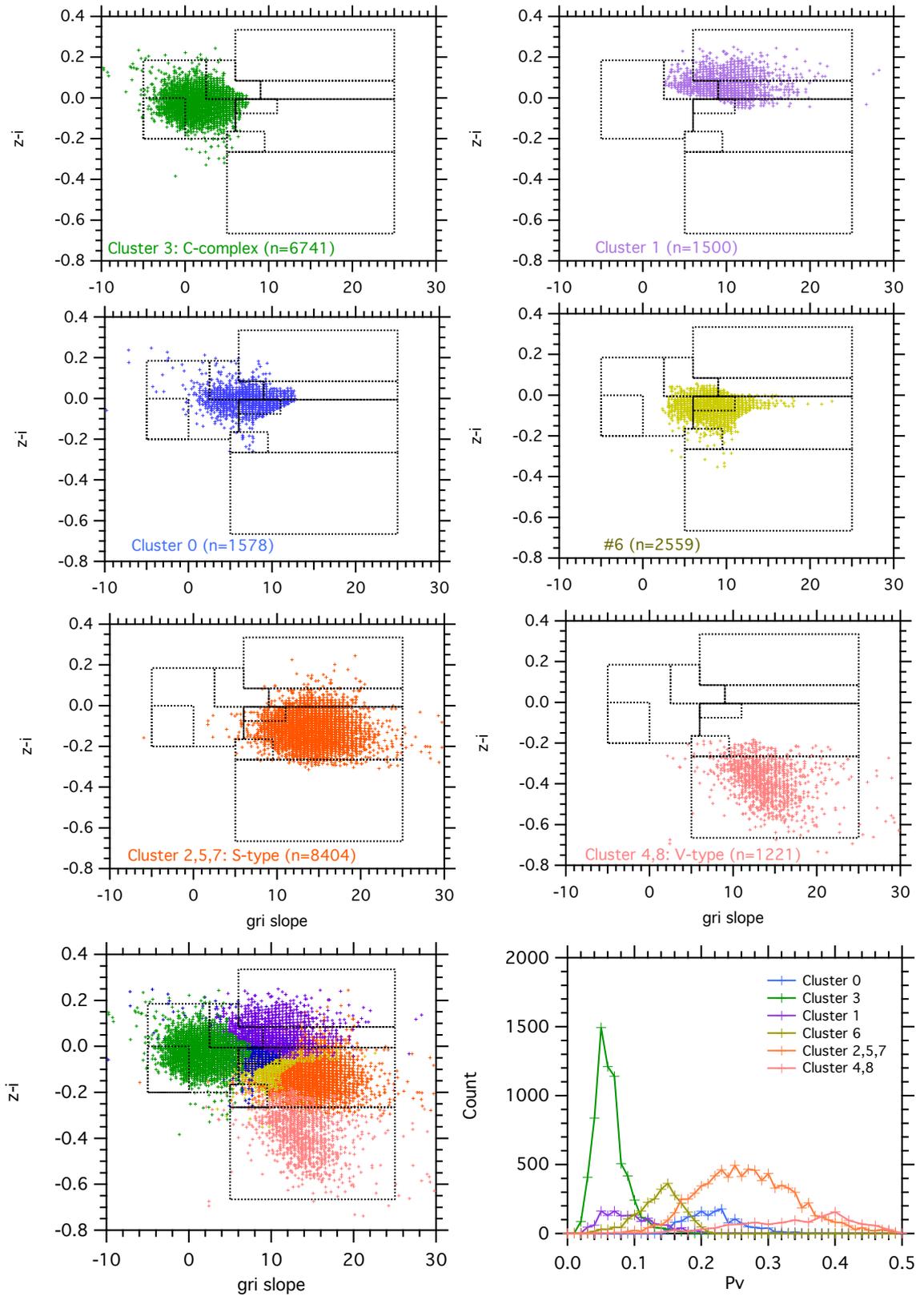

Figure 2: Cluster extracted from the analysis discussed in the text, grouped according to their supposed link to taxonomic types. The distribution of albedo in each cluster

is presented in the lower right graph. The total number of members is also given for each cluster or group of clusters. Gri slope is in %/100 nm.

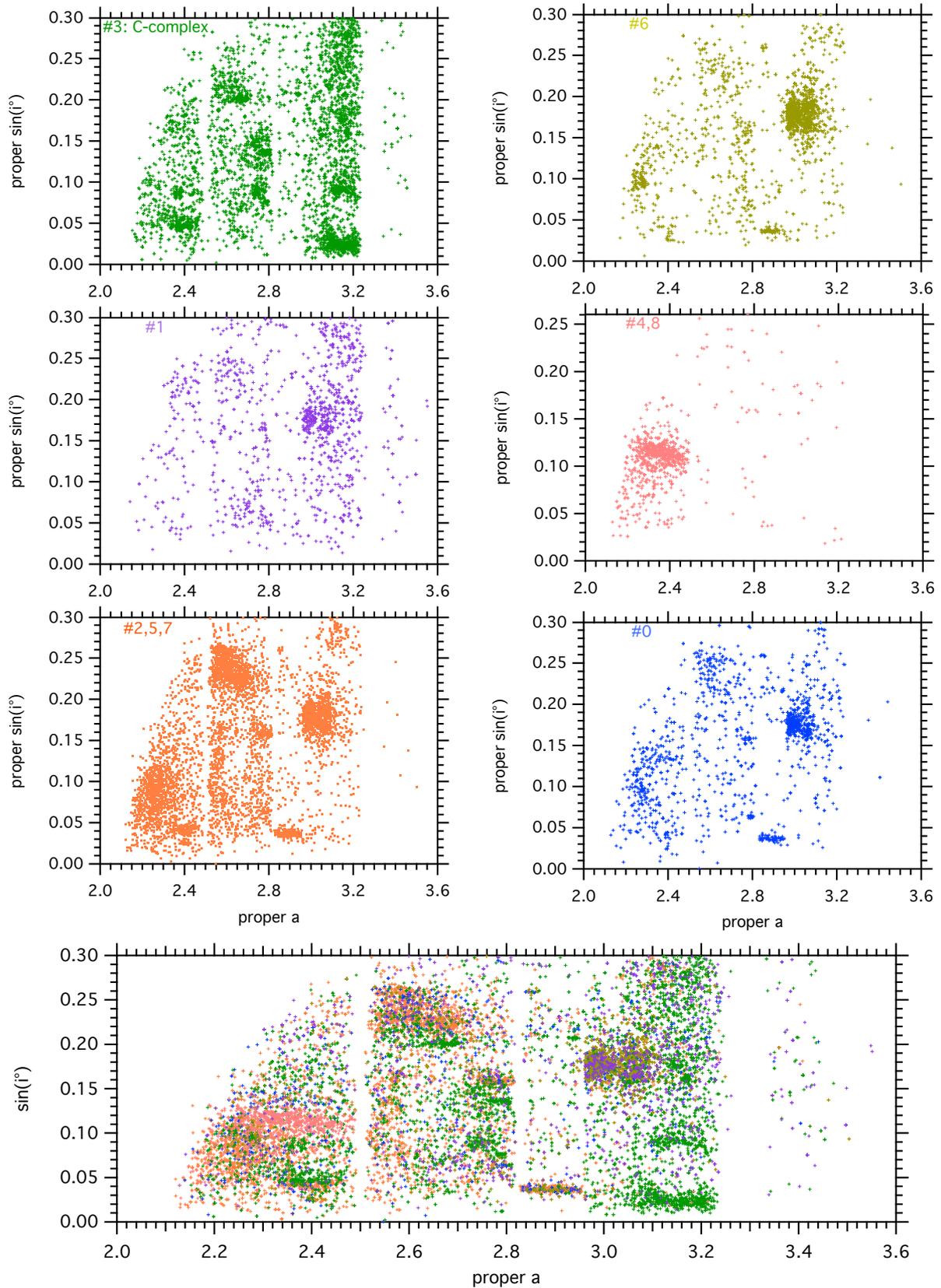

Figure 3: Proper elements (sin(inclinaison) and semi-major axis ) for individual cluster or cluster groups. Only objects with known proper elements in the ADR4 catalogue are shown,

which explains the absence of high-inclination objects in the plots as well as objects with semi-major axis above 3.6.

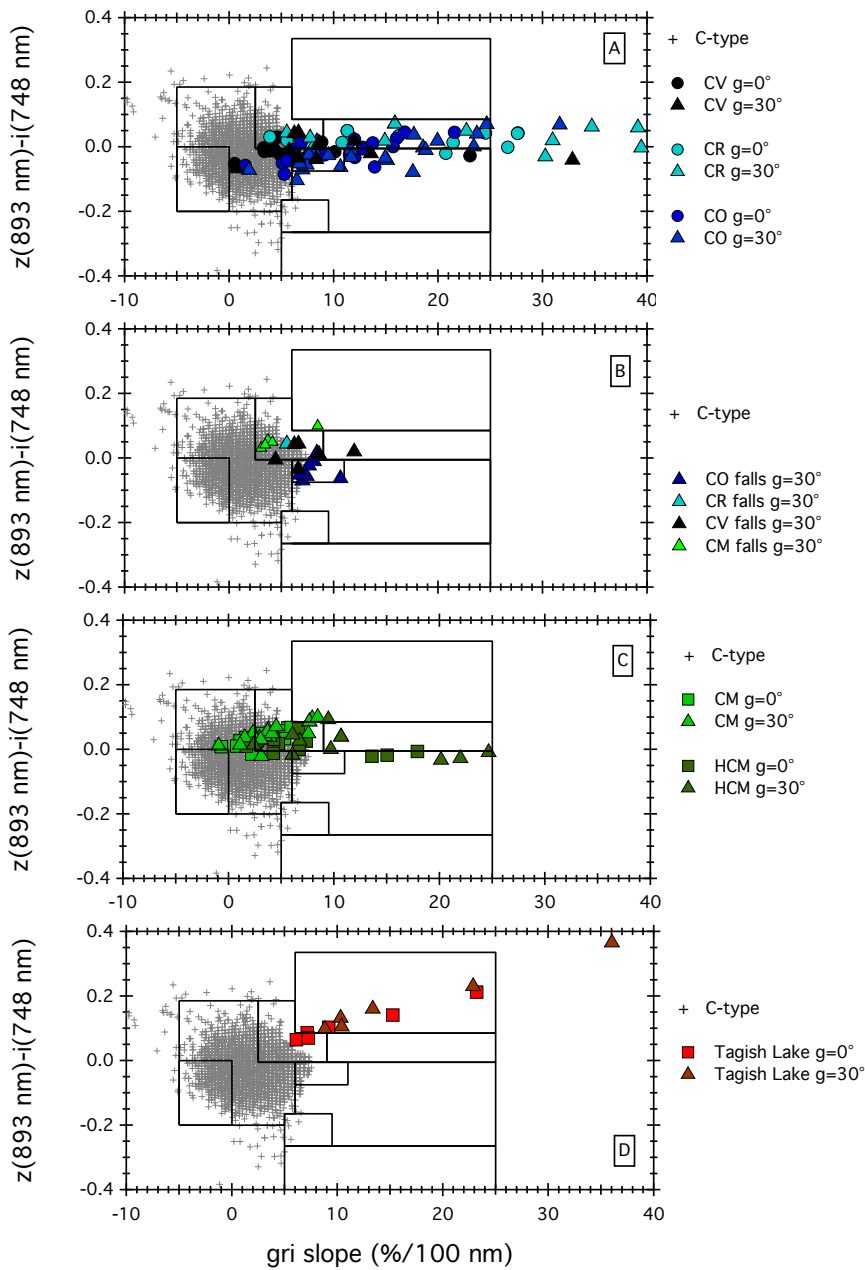

Figure 4: The C-cluster compared to colors of carbonaceous chondrites (meteorite spectra from the RELAB database, except for Tagish Lake measured at IPAG). The boxes correspond to the limits defined by DeMeo and Carry (2013). HCM=heated CM. Gri slope is in %/100 nm. A) CV, CR, CO chondrites falls and find B) CR, CO, CV and CM chondrites falls only C) CM and heated CM falls and finds D) Tagish Lake (ungrouped carbonaceous chondrite fall).

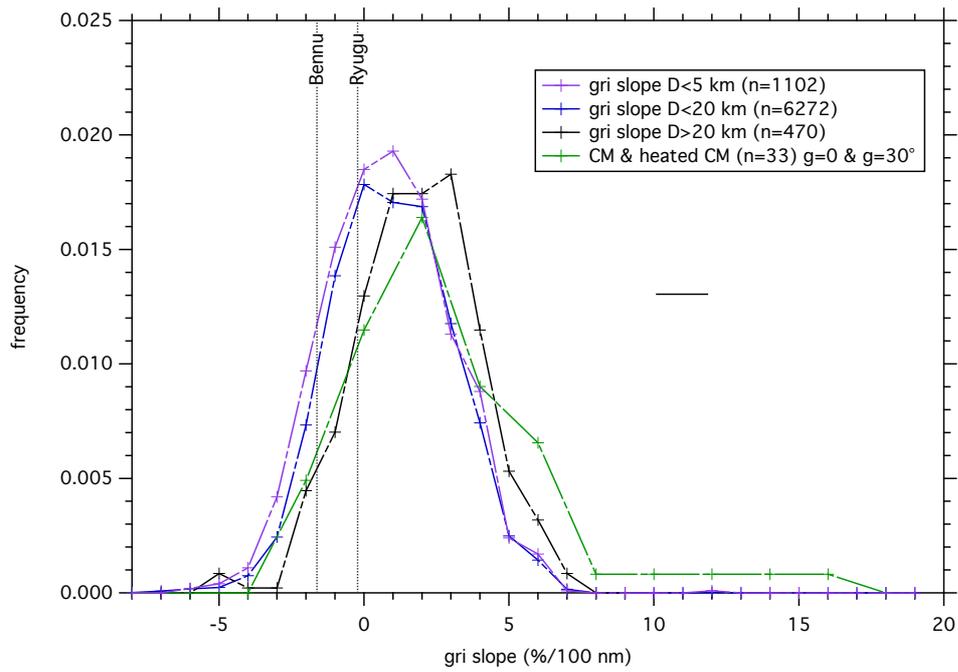

Figure 5: Slope distribution for C-cluster asteroids and CM & heated CM chondrites. The distribution is shown for objects of different sizes. Gri slope is in %/100 nm. The horizontal bar represents the average error on the gri slope.

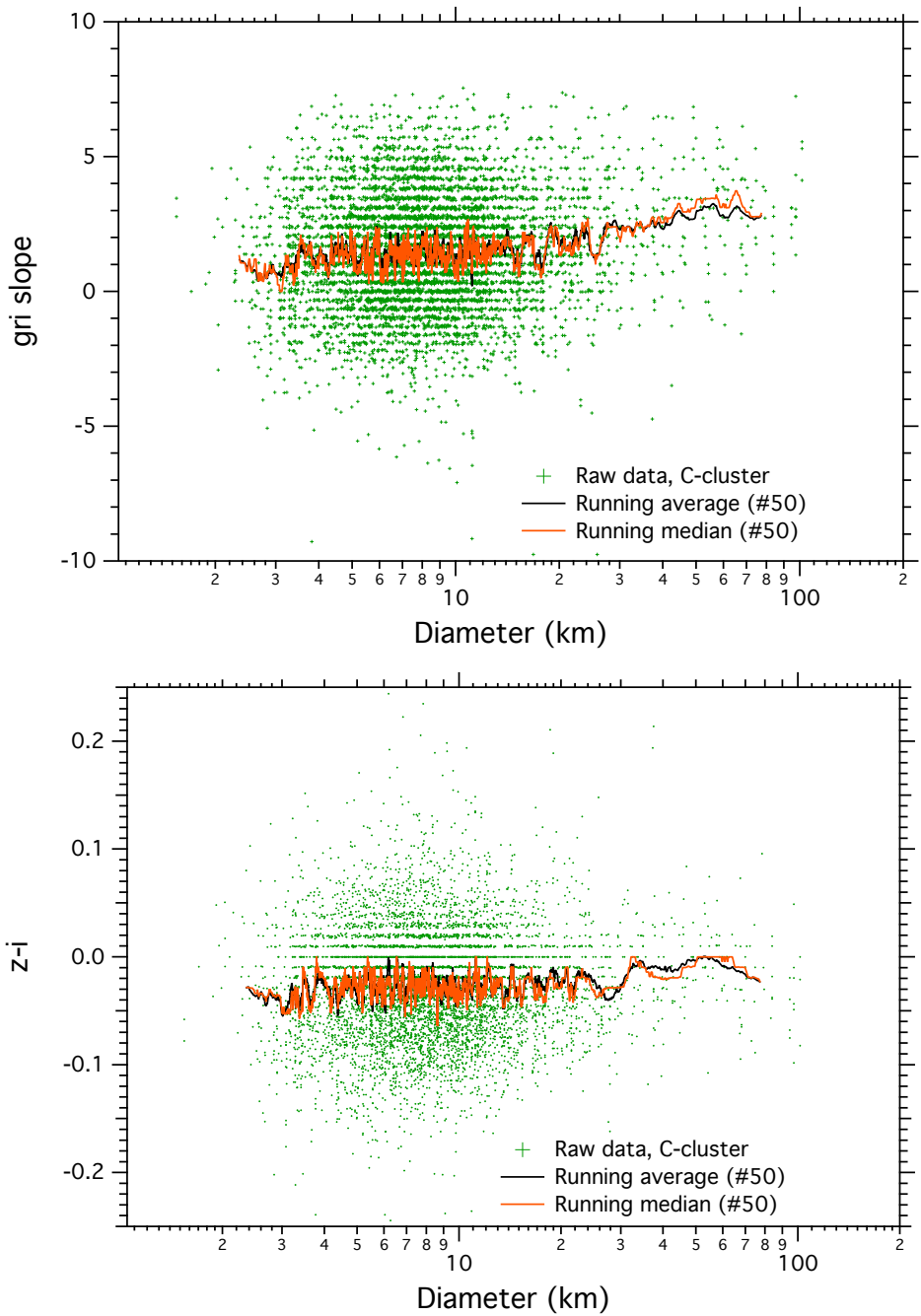

Fig. 6: z-i and gri slope of C-cluster members shown as a function of asteroid diameter. A 50 elements running average and median were calculated and are shown on this graph. Gri slope is in %/100 nm.

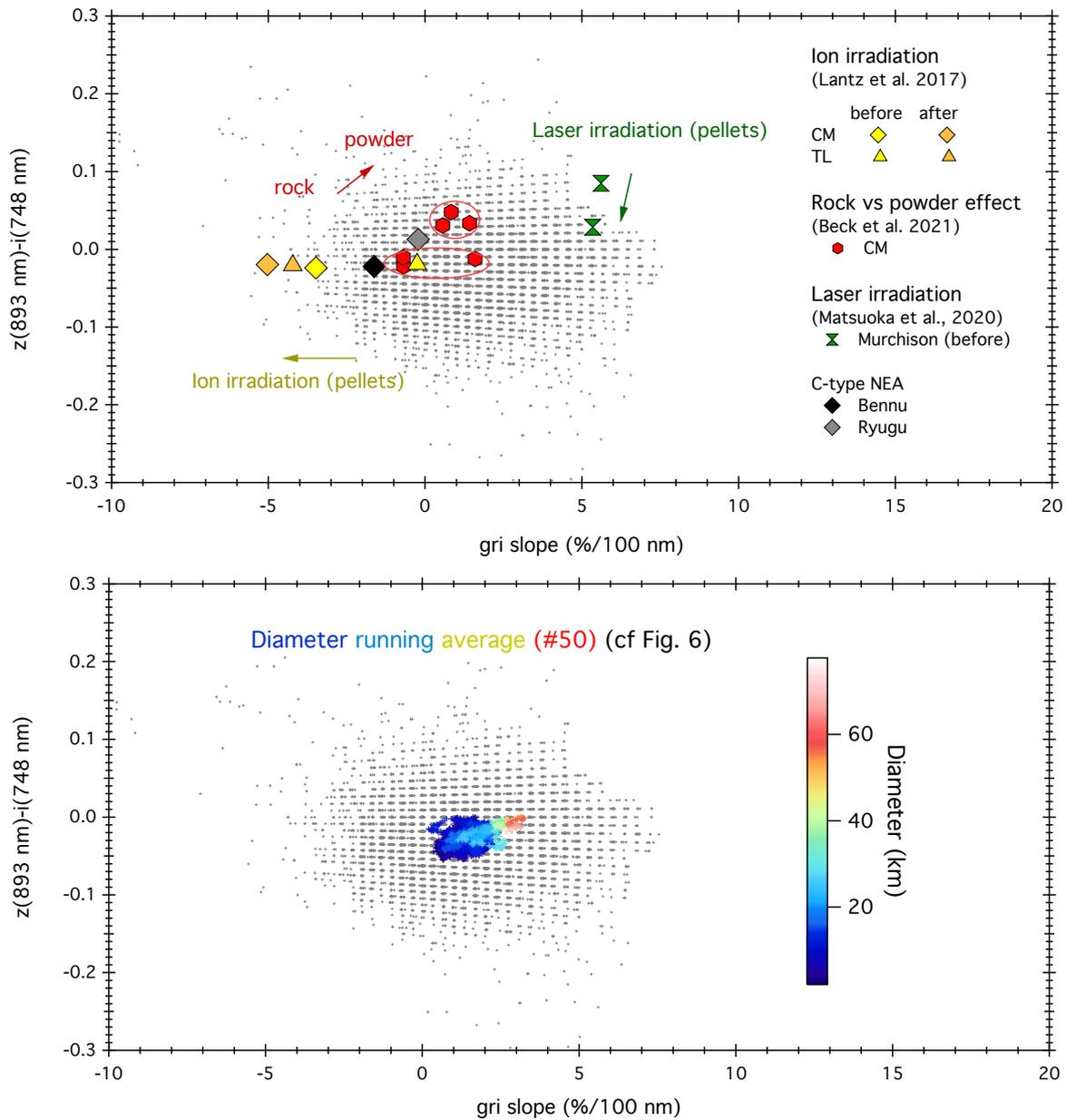

Figure 7: Top) The C-cluster asteroids in the *z-i* vs *gri* slope space and the vector associated to grain size and space weathering simulations. Bottom) *z-i* vs *gri* slope diagram from the running average (see Fig. 6). The data are color coded according to asteroid diameter. This reveals an increasing bluing with decreasing diameter, together with a decrease of *z-i*. This trend is in line with a rock vs powder effect. Gri slope is in %/100 nm.